\documentclass[preprint]{elsarticle} %arXiv

\usepackage{amsfonts}
\usepackage{eurosym}
\usepackage{amssymb,bbold}
\usepackage{amsmath}
\usepackage{natbib}
\usepackage{mathrsfs}
\usepackage{hyperref}
\biboptions{sort&compress}

\begin{document}

\begin{frontmatter}

\title{Effects of quantum deformation on the
  spin-1/2 Aharonov-Bohm problem}

\author[uepg]{F. M. Andrade}
\ead{fmandrade@uepg.br}
\author[ufma]{E. O. Silva}
\ead{edilbertoos@pq.cnpq.br}

\address[uepg]{
  Departamento de Matem\'{a}tica e Estat\'{i}stica,
  Universidade Estadual de Ponta Grossa,
  84030-900 Ponta Grossa-PR, Brazil
}
\address[ufma]{
  Departamento de F\'{i}sica,
  Universidade Federal do Maranh\~{a}o,
  Campus Universit\'{a}rio do Bacanga,
  65085-580 S\~{a}o Lu\'{i}s-MA, Brazil
}

\begin{abstract}
In this letter we study the Aharonov-Bohm problem for a spin-1/2
particle in the quantum deformed framework generated by the
$\kappa$-Poincar\'{e}-Hopf algebra.
We consider the nonrelativistic limit of the $\kappa$-deformed
Dirac equation and use the spin-dependent term to impose an
upper bound on the magnitude of the deformation parameter
$\varepsilon$.
By using the self-adjoint extension approach, we examine the
scattering and bound state scenarios.
After obtaining the scattering phase shift and the $S$-matrix,
the bound states energies are obtained by analyzing the pole
structure of the latter.
Using a recently developed general regularization prescription
[Phys. Rev. D. \textbf{85}, 041701(R) (2012)], the self-adjoint
extension parameter is determined in terms of the physics of the
problem.
For last, we analyze the problem of helicity conservation.
\end{abstract}

\begin{keyword}
%% keywords here, in the form: keyword \sep keyword
$\kappa $-Poincar\'{e}-Hopf algebra \sep self-adjoint extension
\sep Aharonov-Bohm \sep scattering \sep helicity
%% PACS codes here, in the form: \PACS code \sep code

%% MSC codes here, in the form: \MSC code \sep code
%% or \MSC[2008] code \sep code (2000 is the default)

\end{keyword}

\end{frontmatter}

\section{Introduction}
\label{sec:introduction}
Theory of quantum deformations based on the
$\kappa$-Poincar\'{e}-Hopf algebra has been a alternative
framework for studying relativistic and nonrelativistic quantum
systems.
The Hopf-algebraic description of $\kappa$-deformed Poincar\'{e}
symmetries, with $\kappa$ a masslike fundamental deformation
parameter, was introduced in
\cite{PLB.1991.264.331,PLB.1992.293.344}.
In this context, the space-like $\kappa$-deformed Minkowski
spacetime is the more interesting among them because its
phenomenological applications.
Such $\kappa$-deformed Poincar\'{e}-Hopf algebra established in
Refs. \cite{PLB.1991.264.331,PLB.1992.293.344,PLB.1993.302.419,
PLB.1993.318.613,PLB.1994.329.189,PLB.1994.334.348}
is defined by the following commutation relations
\begin{subequations}
\label{eq:algebra}
\begin{equation}
\left[\Pi_{\nu},\Pi_{\mu}\right]=0,
\label{eq:algebraa}
\end{equation}
\begin{equation}
\left[M_{i},\Pi_{\mu}\right]=
(1-\delta_{0\mu})i\epsilon_{ijk}\Pi_{k},
\label{eq:algebrab}
\end{equation}
\begin{equation}
\left[L_{i},\Pi_{\mu}\right]=i [\Pi_{i}]^{\delta_{0\mu}}
[\delta_{ij}\varepsilon^{-1}
\sinh \left( \varepsilon \Pi_{0}\right)]^{1-\delta_{0\mu}},
\label{eq:algebrac}
\end{equation}
\begin{equation}
\left[M_{i},M_{j}\right]=i\epsilon_{ijk} M_{k},\qquad
\left[M_{i},L_{j}\right]=i\epsilon_{ijk} L_{k},
\label{eq:algebrad}
\end{equation}
\begin{equation}
  \left[L_{i},L_{j}\right]=
  -i\epsilon_{ijk}
  \left[
    M_{k}\cosh \left(\varepsilon \Pi_{0}\right)-
    \frac{\varepsilon ^{2}}{4}\Pi_{k}\Pi_{l} M_{l}
  \right],
  \label{eq:algebrae}
\end{equation}
\end{subequations}
where $\varepsilon$ is defined by
\begin{equation}
  \varepsilon=\kappa^{-1}=
  \lim_{R\rightarrow \infty }(R\ln q),
\end{equation}
with $R$ being the de Sitter curvature and $q$ is a real
deformation parameter, $\Pi_{\mu }=(\Pi_{0},\boldsymbol{\Pi})$
are the $\kappa$-deformed generators for energy and momenta.
Also, the $M_{i}$, $L_{i}$ represent the spatial rotations and
deformed boosts generators, respectively.
The coalgebra and antipode for the $\kappa$-deformed
Poincar\'{e} algebra was established in
Ref. \cite{AP.1995.243.90}.

The physical properties of $\kappa$-deformed relativistic
quantum systems can be accessed by solving the $\kappa$-deformed
Dirac equation \cite{PLB.1993.302.419,PLB.1993.318.613,
CQG.2004.21.2179,JHEP.2004.2004.28}.
The deformation parameter $\kappa$ can be usually interpreted as
being the Planck mass $M_{P}$ \cite{PLB.2012.711.122}.
The $\kappa$-deformation has implications for various properties
of physical systems as for example, vacuum energy divergent
\cite{PRD.2007.76.125005}, Landau levels \cite{PLB.1994.339.87},
spin-1/2 Aharonov-Bohm (AB) interaction creating additional
bound states \cite{PLB.1995.359.339}, Dirac oscillator
\cite{EPL.1997.39.583}, Dirac-Coulomb problem
\cite{PLB.1993.318.613} and constant magnetic interaction
\cite{MPLA.1995.10.1969}.
In Ref. \cite{PLB.1995.359.339} the spin-1/2 AB problem was
solved for the first time in connection with the theory of
quantum deformations. 
The AB problem \cite{PR.1959.115.485} has been extensively
studied in different contexts in recent years 
\cite{PRD.2011.84.045002,PRL.2012.108.230404,JPA.2010.43.075202,
PRD.2012.85.041701, PRA.2012.86.040101,PRL.2012.108.153901,
PRD.2011.83.125025,PRD.2012.86.125015}.
In this letter we study the scattering scenario of the model
addressed in Ref. \cite{PLB.1995.359.339} where only the bound
state problem was considered.
We solve the problem by following the self-adjoint extension
approach
\cite{CMP.1991.139.103,JMP.1985.26.2520,Book.2004.Albeverio}
and by using the general regularization prescription proposed in
\cite{PRD.2012.85.041701} we determine the self-adjoint
extension parameter in terms of the physics of the problem.
Such procedure allows discuss the problem of helicity
conservation and, as a alternative approach, we obtain the bound
states energy from the poles of $S$-matrix.

The plan of our Letter is the following.
In Section \ref{sec:k-deformed-SPE} we introduce the
$\kappa$-deformed Dirac equation to be solved and take its
nonrelativistic limit in order to study the physical
implications of $\kappa$-deformation in the spin-1/2 AB
problem.
A new contribution to the nonrelativistic Hamiltonian arises in
this approach. 
These new term imply a direct correction on the anomalous
magnetic moment term. 
We impose a upper bound on the magnitude of the deformation
parameter $\varepsilon$.
The Section \ref{sec:selfae} is devoted to study the
$\kappa$-deformed Hamiltonian via self-adjoint extension
approach and presented some important properties of the
$\kappa$-deformed wave function.
In Section \ref{sec:scatt-bound} are addressed the scattering and
bound states scenario within the framework of $\kappa$-deformed
Schr\"{o}dinger-Pauli equation.
Expressions for the phase shift, $S$-matrix, and bound states
are derived.
We also derive a relation between the self-adjoint extension
parameter and the physical parameters of the problem.
For last, we make a detailed analysis of the helicity
conservation problem in the present framework.
A brief conclusion in outlined in Section \ref{sec:conclusion}.

\section{$\kappa$-deformed Schr\"{o}dinger-Pauli
  equation}
\label{sec:k-deformed-SPE}

In the minimal coupling prescription the (3+1)-dimensional
$\kappa$-deformed Dirac equation supported by the algebra in
Eq. \eqref{eq:algebra} up to $O(\varepsilon)$ order was derived
in Ref. \cite{PLB.1995.359.339} (see also Refs. therein).
We here analyze the (2+1)-dimensional $\kappa$-deformed Dirac
equation, which follows from the decoupling of (3+1)-dimensional
$\kappa$-deformed Dirac equation for the specialized case where
$\partial_3=0$ and $A_{3}=0$, into two uncoupled two-component
equations, such as implemented in
Refs. \cite{PRD.1978.18.2932,NPB.1988.307.909,PRL.1989.62.1071}.
This way, the planar $\kappa$-deformed Dirac equation
($\hbar=c=1$) is
\begin{equation}
  \hat{H}\psi=
  \left[
    \beta \boldsymbol{\gamma} \cdot \boldsymbol{\Pi}
    +\beta M +
    \frac{\varepsilon}{2}
    \left(
      M \boldsymbol{\gamma} \cdot \boldsymbol{\Pi} +
      e s \boldsymbol{\sigma} \cdot \boldsymbol{B}
    \right)
  \right]\psi= \overline{E}\psi,
\label{eq:defdirac}
\end{equation}
where $\psi$ is a two-component spinor,
$\boldsymbol{\Pi}=\boldsymbol{p}-e\boldsymbol{A}$ is the
generalized momentum, and $s$ is twice the spin value, with
$s=+1$ for spin ``up'' and $s=-1$ for spin ``down''.
The $\gamma$-matrices in $(2+1)$ are given in terms of the Pauli
matrices
\begin{equation}
  \beta=\gamma_{0}=\sigma_{3}, \qquad
  \gamma_{1}=i\sigma_{2}, \qquad
  \gamma_{2}=-is\sigma_{1}.
\end{equation}
Here few comments are in order.
First, the $\kappa$-deformed Dirac equation is defined in the
commutative spacetime and the corresponding  $\gamma$-matrices
are independent of the deformation parameter $\kappa$
\cite{MPLA.2011.26.1103}.
Second, it is important to observe that in
Ref. \cite{PLB.1995.359.339} the authors only consider the
negative value of the spin projection, here our approach considers
a more general situation.

We shall now take the nonrelativistic limit of
Eq. \eqref{eq:defdirac}.
Writing $\psi=(\chi,\phi)^{T}$, where $\chi$ and $\phi$ are the
``large'' and ``small'' components of the spinor, and
using $\overline{E}=M+E$ with $M\gg E$, after expressing the
lower component $\phi$ in terms of the upper one, $\chi$, we get
the  $\kappa$-deformed Schr\"{o}dinger-Pauli equation for the
large component
\begin{equation}
H\chi=E\chi,
\end{equation}
with
\begin{equation}
  H=\frac{1}{2M}
  \left[
    \Pi_{1}^{2}+\Pi_{2}^{2}-(1-M\varepsilon) e s B_{3}
  \right],
\label{eq:nrdefdirac}
\end{equation}
where it was assumed that $\varepsilon^{2}\cong 0$.
It can be seen from \eqref{eq:nrdefdirac} that the magnetic
moment has modified by a quantity proportional to the deformation
parameter.

Another effect enclosed in Hamiltonian \eqref{eq:nrdefdirac} is
concerned with the anomalous magnetic moment of the electron.
The electron  magnetic moment is 
$\boldsymbol{\mu} =-\mu \boldsymbol{\sigma}$, with $\mu=e/2M$,
and $g=2$ the gyromagnetic factor.
The anomalous magnetic moment of the electron is given by
$g=2(1+a)$, with $a=\alpha /2\pi=0.00115965218279$ representing the
deviation in relation to the usual case \cite{RMP.2012.84.1527}.
In this case, the magnetic interaction is
$\bar{H}=\mu(1+a)(\boldsymbol{\sigma}\cdot\boldsymbol{B})$.
In accordance with very precise measurements and quantum
electrodynamics (QED) calculations \cite{PRL.2006.97.030802}, precision corrections to
this factor are now evaluated at the level of $1$ part in
$10^{11}$, that is, $\Delta a\leq 3\times 10^{-11}$.
In our case, the Hamiltonian \eqref{eq:nrdefdirac} provides
$\kappa$-tree-level contributions to the usual $g=2$
gyromagnetic factor, which can not be larger than $a=0.00116$
(the current experimental value for the anomalous magnetic
moment).
The total $\kappa$-deformed magnetic interaction in
Eq. \eqref{eq:nrdefdirac} is
\begin{equation}
  H_{\text{magn}}=(1-M\varepsilon) s(\boldsymbol{\mu \cdot B}).
  \label{eq:hmagn}
\end{equation}
For the magnetic field along the $z$-axis and a spin-polarized
configuration in the $z$-axis, this interaction assumes the form
\begin{equation}
\left( 1-M\varepsilon \right) s\mu B_{z},
\end{equation}
with $M\varepsilon$ representing the $\kappa$-tree-level
correction that should be smaller than $0.00116$.
Under such consideration, we obtain the following upper bound
for $\varepsilon $:
\begin{equation}
\varepsilon <2.27\times 10^{-9}\;(eV) ^{-1},
\label{eq:bound}
\end{equation}
where we have used $M=5.11\times 10^{5}eV$.

We now pass to study the $\kappa$-deformed Schr\"{o}dinger-Pauli
equation in the AB background potential
\cite{PR.1959.115.485}.
The vector potential of the AB interaction, in the
Coulomb gauge, is
\begin{equation}
  \mathbf{A}=-
  \frac{\alpha}{r}
  \hat{\boldsymbol{\varphi}},
  \qquad A_{0}=0,
  \label{eq:vectora}
\end{equation}
where $\alpha = \Phi /\Phi_{0}$ is the flux parameter with
$\Phi_{0}=2\pi/e$.
The magnetic field is given in the usual way
\begin{equation}
 e \mathbf{B}=e\nabla \times \mathbf{A}=
  -\alpha  \frac{\delta (r)}{r}\mathbf{\hat{z}}.
\label{eq:vectorb}
\end{equation}
So, the  $\kappa$-deformed Schr\"{o}dinger-Pauli equation can be
written as
\begin{equation}
  \frac{1}{2M}
  \left[
    H_{0}+{\eta}
  \frac{\delta(r)}{r}
  \right]
  \chi = E\chi,
  \label{eq:schor}
\end{equation}
with
\begin{equation}
  H_{0}=
  \left(
    \frac{1}{i}\boldsymbol{\nabla}
    - e \mathbf{A}
  \right)^{2},
\end{equation}
and
\begin{equation}
  \label{eq:dcoup}
  \eta=(1-M\varepsilon)\alpha s,
\end{equation}
is the coupling constant of the $\delta(r)/r$ potential.

For the present system the total angular momentum operator in
the $z$ direction,
\begin{equation}
\hat{J}_{3}=-i\partial_{\varphi}+\frac{1}{2}\sigma_{3},
\end{equation}
commutes with the effective Hamiltonian.
So, it is possible to express the eigenfunctions of the two
dimensional Hamiltonian in terms of the eigenfunctions of
$\hat{J}_{3}$.
The eigenfunctions of this operator are
\begin{equation}
  \psi=
  \left(
    \begin{array}{c}
      \chi \\
      \phi
    \end{array}
  \right)=
  \left(
    \begin{array}{c}
      f_{m}(r)\; e^{i(m_{j}-1/2)\varphi} \\
      g_{m}(r)\; e^{i(m_{j}+1/2)\varphi}
    \end{array}
  \right),
  \label{eq:wavef}
\end{equation}
with $m_{j}=m+1/2=\pm 1/2,\pm 3/2, \ldots$, with
$m\in\mathbb{Z}$.
Inserting this into Eq. \eqref{eq:schor}, we can extract
the radial equation for $f_{m}(r)$ ($k^{2}= 2 M E$)
\begin{equation}
h f_{m}(r)=k^{2} f_{m}(r),
 \label{eq:eigen}
\end{equation}
where
\begin{equation}
h=h_{0}+\eta \frac{\delta(r)}{r},
\label{eq:hfull}
\end{equation}
\begin{equation}
  h_{0}=
  -\frac{d^{2}}{dr^{2}}
  -\frac{1}{r}\frac{d}{dr}
  +\frac{(m+\alpha)^{2}}{r^{2}}.
  \label{eq:hzero}
\end{equation}
The Hamiltonian in Eq. \eqref{eq:hfull} is singular at
the origin.
This problem can then be treated by the method of the
self-adjoint extension \cite{Book.2004.Albeverio}, which we pass
to discuss in the next Section.

\section{Self-adjoint extension analysis}
\label{sec:selfae}

The operator $h_{0}$, with domain $\mathcal{D}(h_{0})$, is
self-adjoint if $h_{0}^{\dagger}=h_{0}$ and
$\mathcal{D}(h_{0}^{\dagger})=\mathcal{D}(h_{0})$.
For smooth functions, $\xi \in C_{0}^{\infty}(\mathbb{R}^2)$ with
$\xi(0)=0$, we should have $h \xi =h_{0} \xi$, and hence it is
reasonable to interpret the Hamiltonian \eqref{eq:hfull} as a
self-adjoint extension of $h_{0}|_{C_{0}^{\infty}(\mathbb{R}^{2}/\{0\})}$
\cite{crll.1987.380.87,JMP.1998.39.47,LMP.1998.43.43}.
In order to proceed to the self-adjoint extensions of
\eqref{eq:hzero}, we decompose the Hilbert space
$\mathfrak{H}=L^{2}(\mathbb{R}^{2})$
with respect to the angular momentum
$\mathfrak{H}=\mathfrak{H}_{r}\otimes\mathfrak{H}_{\varphi}$, where
$\mathfrak{H}_{r}=L^{2}(\mathbb{R}^{+},rdr)$ and
$\mathfrak{H}_{\varphi}=L^{2}(\mathcal{S}^{1},d\varphi)$,
with $\mathcal{S}^{1}$ denoting the unit sphere in
$\mathbb{R}^{2}$.
The operator $-\partial^{2}/\partial\varphi^{2}$ is
essentially self-adjoint in
$L^{2}(\mathcal{S}^{1},d\varphi)$
\cite{Book.1975.Reed.II} and we obtain the operator $h_{0}$ in
each angular momentum sector.
Now, using the unitary operator
$U: L^{2}(\mathbb{R}^{+},rdr)
\to L^{2}(\mathbb{R}^{+}, dr)$,
given by $(U \xi)(r)=r^{1/2}\xi(r)$, the operator $h_{0}$ becomes
\begin{equation}
  \tilde{h}_{0}=
  U h_{0} U^{-1}=
  -\frac{d^{2}}{dr^{2}}
  -\left[
    (m+\alpha)^{2}-
    \frac{1}{4}
  \right]
  \frac{1}{r^{2}},
\end{equation}
which is essentially self-adjoint for $|m+\alpha| \geq 1$, while
for $|m+\alpha|< 1$ it admits a one-parameter family of
self-adjoint extensions \cite{Book.1975.Reed.II},
$h_{0,\lambda_{m}}$, where $\lambda_{m}$ is the self-adjoint
extension parameter.
To characterize this family we will use the approach in
\cite{JMP.1985.26.2520}, which is based in a boundary
conditions at the origin.

Following the approach in
Refs. \cite{JMP.1985.26.2520,Book.2004.Albeverio}, all the
self-adjoint extensions $h_{0,\lambda_{m}}$  of $h_{0}$ are
parametrized by the boundary condition at the origin
\begin{equation}
\label{eq:bc}
  f_{0,\lambda_{m}}=\lambda_{m} f_{1,\lambda_{m}},
\end{equation}
with
\begin{align}
  f_{0,\lambda_{m}}&=\lim_{r\rightarrow 0^{+}}r^{|m+\alpha|}f_{m}(r),
  \\
  f_{1,\lambda_{m}}&=\lim_{r\rightarrow 0^{+}}\frac{1}{r^{|m+\alpha|}}
  \left[
    f_{m}(r)-f_{0,\lambda_{m}}\frac{1}{r^{|m+\alpha|}}
  \right],
\end{align}
where $\lambda_{m}\in \mathbb{R}$ is the self-adjoint extension
parameter.
The self-adjoint extension parameter $\lambda_{m}$ has a
physical interpretation, it represents the scattering length
\cite{Book.2011.Sakurai} of $h_{0,\lambda_{m}}$
\cite{Book.2004.Albeverio}.
For $\lambda_{m}=0$ we have the free Hamiltonian (without the
$\delta$ function)  with regular wave functions at the origin,
and for $\lambda_{m}\neq 0$  the boundary condition in
Eq. \eqref{eq:bc}
permit a $r^{-|m+\alpha|}$ singularity in the wave functions at
the origin.

\section{Scattering and bound states analysis}
\label{sec:scatt-bound}

The general solution for Eq. \eqref{eq:eigen} in the $r\neq 0$
region can be written as
\begin{equation}
\label{eq:sol1}
f_{m}(r)=a_{m}J_{|m+\alpha|}(kr)+b_{m}Y_{|m+\alpha|}(kr),
\end{equation}
with $a_{m}$ and $b_{m}$ being constants and $J_{\nu}(z)$ and
$Y_{\nu}(z)$ are the Bessel functions of first and second kind,
respectively.
Upon replacing $f_{m}(r)$ in the boundary condition
\eqref{eq:bc},
 we obtain
\begin{align}
  \lambda_{m} \, a_{m} \, \upsilon  \, k^{|m+\alpha|}
  = {} &
  b_{m}\left[\zeta k^{-|m+\alpha|}\right.
   \nonumber \\
  & \left. -\lambda_{m} \big(\beta  k^{|m+\alpha|} +
   \zeta \nu k^{-|m+\alpha|}
  \lim_{r\rightarrow 0^{+}}r^{2-2|m+\alpha|}\big)\right],
  \label{eq:bcf}
\end{align}
where
\begin{align}
\upsilon & = \frac{1}{2^{|m+\alpha|}\Gamma(1+|m+\alpha|)}, &
\zeta & =-\frac{2^{|m+\alpha|}\Gamma(|m+\alpha|)}{\pi},
\nonumber  \\
\beta & =-\frac{\cos (\pi |m+\alpha|)
 \Gamma(-|m+\alpha|)}{\pi 2^{|m+\alpha|}},&
\nu &   =\frac{k^{2}}{4(1-|m+\alpha|)}.
\end{align}
In Eq. \eqref{eq:bcf}, $\lim_{r\rightarrow 0^{+}}r^{2-2|m+\alpha|}$
is divergent if $|m+\alpha|\geq 1$, hence $b_{m}$ must be zero.
On the other hand, $\lim_{r\rightarrow 0^{+}}r^{2-2|m+\alpha|}$
is finite for $|m+\alpha|<1$, it means that there arises the
contribution of the irregular solution $Y_{|m+\alpha|}(kr)$.
Here, the presence of an irregular solution contributing to the
wave function stems from the fact the Hamiltonian $h$ is not a
self-adjoint operator when $|m+\alpha|<1$ (cf., Section
\ref{sec:selfae}), hence such irregular solution must be
associated with a self-adjoint extension of the operator $h_{0}$
\cite{JPA.1995.28.2359,PRA.1992.46.6052}.
Thus, for $|m+\alpha|<1$, we have
\begin{equation}
  \lambda_{m}a_{m}\upsilon k^{|m+\alpha|}=
  b_{m}(\zeta k^{-|m+\alpha|}-
  \lambda_{m}\beta k^{|m+\alpha|}),
\end{equation}
and by substituting the values of $\upsilon$, $\zeta$ and $\beta$
into above expression we find
\begin{equation}
  b_{m}=-\mu_{m}^{\lambda_{m}}a_{m},
\end{equation}
where
\begin{equation}
  \mu_{m}^{\lambda_{m}}=
  \frac
  {\lambda_{m}k^{2|m+\alpha|}
    \Gamma(1-|m+\alpha|)\sin(\pi |m+\alpha| )}
  {B_{k}},
  \label{eq:mul}
\end{equation}
and
\begin{align}
  B_{k}= {} & \lambda_{m}k^{2|m+\alpha|}
  \Gamma{(1-|m+\alpha|)}\cos (\pi|m+\alpha|)\nonumber \\
  & + 4^{|m+\alpha|}\Gamma(1+|m+\alpha|).
\end{align}
Since a $\delta$ function is a very short range potential, it
follows that the asymptotic behavior of $f_{m}(r)$ for
$r\rightarrow \infty$ is given by \cite{JPA.2010.43.354011}
\begin{equation}
  f_{m}(r)\sim \sqrt{\frac{2}{\pi kr}}
  \cos \left( kr-\frac{|m|\pi}{2}-
  \frac{\pi}{4}+
  \delta_{m}^{{\lambda_{m}}}(k,\alpha)\right) ,
\label{eq:f1asim}
\end{equation}
where $\delta_{m}^{{\lambda_{m}}}(k,\alpha)$  is a
scattering phase shift.
The phase shift is a measure of the argument difference to the
asymptotic behavior of the solution $J_{|m|}(kr)$ of the radial
free equation which is regular at the origin.
By using the asymptotic behavior of the Bessel functions
\cite{Book.1972.Abramowitz} into Eq. \eqref{eq:sol1} we obtain
\begin{align}
  \label{eq:scattsol}
  f_{m}(r)
  \sim {} &
  a_{m}\sqrt{\frac{2}{\pi kr}}
  \left[
    \cos\left(kr-\frac{\pi |m+\alpha|}{2}-\frac{\pi}{4}\right)
  \right. \nonumber\\
  &
  \left.
    - \mu_{m}^{{\lambda_{m}}}
    \sin \left( kr-\frac{\pi |m+\alpha|}{2}-\frac{\pi}{4}\right)
  \right] .
\end{align}
By comparing the above expression with Eq. \eqref{eq:f1asim}, we
found
\begin{equation}
\delta_{m}^{{\lambda_{m}}}(k,\alpha)=
\Delta_{m}^{AB}(\alpha)+\theta_{{\lambda_{m}}},
\label{eq:phaseshift}
\end{equation}
where
\begin{equation}
\Delta_{m}^{AB}(\alpha)=\frac{\pi}{2}(|m|-|m+\alpha|),
\end{equation}
is the usual phase shift of the AB scattering and
\begin{equation}
  \theta_{{\lambda_{m}}}=\arctan {(\mu_{m}^{\lambda_{m}})}.
\end{equation}
Therefore, the scattering operator $S_{\alpha,m}^{\lambda_{m}}$
($S$-matrix) for the self-adjoint extension is
\begin{equation}
S_{\alpha,m}^{\lambda_{m}}=
e^{2i\delta_{m}^{{\lambda_{m}}}(k,\alpha)}=
e^{2i\Delta_{m}^{AB}(\alpha)}
  \left[
    \frac{1+i\mu_{m}^{\lambda_{m}}}{1-i\mu_{m}^{\lambda_{m}}}
  \right].
\end{equation}
Using Eq. \eqref{eq:mul}, we have
\begin{align}
  S_{\alpha,m}^{\lambda_{m}}
  = {} &
  e^{2i\Delta_{m}^{AB}(\alpha)} \nonumber \\
  {} &  \times
  \bigg[
  \frac
  {B_{k}+i\lambda_{m}k^{2|m+\alpha|}
    \Gamma(1-|m+\alpha|)\sin(\pi |m+\alpha|)}
  {B_{k}-i\lambda_{m}k^{2|m+\alpha|}
    \Gamma(1-|m+\alpha|)\sin(\pi |m+\alpha|)}
  \bigg].
 \label{eq:smatrix}
\end{align}
Hence, for any value of the self-adjoint extension parameter
$\lambda_{m}$, there is an additional scattering.
If ${\lambda_{m}}=0$, we achieve the corresponding result
for the usual AB problem with Dirichlet boundary condition; in
this case, we recover the expression for the scattering matrix
found in Ref. \cite{AP.1983.146.1},
$S_{\alpha,m}^{\lambda_{m}}=e^{2i\Delta_{m}^{AB}(\alpha)}$.
If we make ${\lambda_{m}}=\infty $, we get
$S_{\alpha,m}^{\lambda_{m}}=e^{2i\Delta_{m}^{AB}(\alpha)+2i\pi |m+\alpha|}$.

In accordance with the general theory of scattering, the poles
of the $S$-matrix in the upper half of the complex plane
\cite{PRC.1999.60.34308} determine the positions of the bound
states in the energy scale.
These poles occur in the denominator of \eqref{eq:smatrix} with
the replacement $k\rightarrow i\kappa$,
\begin{equation}
  B_{i \kappa}+i\lambda_{m}(i \kappa)^{2|m+\alpha|}
  \Gamma(1-|m+\alpha|)\sin(\pi |m+\alpha|)=0.
\end{equation}
Solving the above equation for $E$, we found the bound state
energy
\begin{equation}
  E=-\frac{2}{M}
  \left[-\frac{1}{\lambda_{m}}
    \frac{\Gamma(1+|m+\alpha|)}{\Gamma(1-|m+\alpha|)}
  \right]^{1/|m+\alpha|},
  \label{eq:energy_BG-sc}
\end{equation}
for $\lambda_{m}<0$.
Hence, the poles of the scattering matrix only occur for
negative values of the self-adjoint extension parameter.
In this latter case, the scattering operator can be expressed in
terms of the bound state energy
\begin{equation}
  S_{\alpha,m}^{\lambda_{m}}=e^{2i\Delta_{m}^{AB}(\alpha)}
  \left[
    \frac
    {e^{2 i \pi |m+\alpha|}-(\kappa/k)^{2|m+\alpha|}}
    {1-(\kappa/k)^{2|m+\alpha|}}
  \right].
\end{equation}

The scattering amplitude $f_{\alpha}(k,\varphi)$
can be obtained using the standard methods of scattering theory,
namely
\begin{align}
    f_{\alpha}(k,\varphi)
    = {} &
    \frac{1}{\sqrt{2\pi i k}}\sum_{m=-\infty}^{\infty}
    \left(
      e^{2 i\delta_{m}^{{\lambda_{m}}}(k,\alpha)}-1
    \right)
    e^{im\varphi}   \nonumber \\
    ={} &
    \frac{1}{\sqrt{2\pi i k}}\sum_{m=-\infty}^{\infty}
    \left( e^{2 i \Delta_{m}(\alpha)}
      \left[
        \frac
        {1+ i\mu_{m}^{\lambda_{m}}}
        {1-i\mu_{m}^{\lambda_{m}}}
      \right]
      -1
    \right) e^{im\varphi}.
    \label{eq:scattamp}
\end{align}
In the above equation we can see that the scattering amplitude
differ from the usual AB scattering amplitude off a thin
solenoid because it is energy dependent (cf., Eq. \eqref{eq:mul}).
The only length scale in the nonrelativistic problem is set by
$1/k$, so it follows that the scattering amplitude
would be a function of the angle alone, multiplied by $1/k$
\cite{PRD.1977.16.1815}.
This statement is the manifestation of the helicity conservation
\cite{Book.1967.Sakurai}.
So, one would to expect the commutator of the Hamiltonian with
the helicity operator,
$\hat{h}=\boldsymbol{\Sigma} \cdot \boldsymbol{\Pi}$, to be
zero.
However, when calculated, one finds that
\begin{equation}
  [\hat{H},\hat{h}]=e  \varepsilon
  \left(
    \begin{array}{cc}
      0 & (\boldsymbol{\sigma} \cdot \boldsymbol{B})
      (\boldsymbol{\sigma} \cdot \boldsymbol{\Pi}) \\
      (\boldsymbol{\sigma} \cdot \boldsymbol{B})
      (\boldsymbol{\sigma} \cdot \boldsymbol{\Pi}) & 0
    \end{array}
  \right),
\end{equation}
which is nonzero for $\varepsilon \neq 0$.
So, the inevitable failure of helicity conservation expressed in
Eq. \eqref{eq:scattamp} follow directly from the deformation
parameter $\varepsilon$ and it must be related  with the
self-adjoint extension parameter, because the scattering
amplitude depend on $\lambda_{m}$.
Indeed, as it was shown in \cite{PRD.2012.85.041701} it is
possible to find a relation between the self-adjoint extension
parameter and the coupling constant $\eta$ in \eqref{eq:dcoup}.
By direct inspection we can claim that such relation is
\begin{equation}
  \frac{1}{\lambda_{m}}=
  -\frac{1}{r_{0}^{2|m+\alpha|}}
  \left(
    \frac
    {\eta + |m+\alpha|}
    {\eta - |m+\alpha|}
  \right),
  \label{eq:lambdaj}
\end{equation}
where $r_0$ is a very small radius smaller than the Compton wave
length  $\lambda_C$ of the electron \cite{PLB.1994.333.238},
which comes from the regularization of the $\delta$ function
(for detailed analysis see \cite{arXiv.quant-ph.1207.0214}).
The above relation is only valid for $\lambda_{m}<0$
(when we have scattering and bound states), consequently we
have $|\eta|\geq |m+\alpha|$ and due to $|m+\alpha|<1$ it is
sufficient to consider $|\eta|\geq 1$ to guarantee $\lambda_{m}$
to be negative.
A necessary condition for a $\delta$ function generates an
attractive potential, which is able to support bound states, is
that the coupling constant must be negative.
Thus, the existence of bound states requires
\begin{equation}
\eta \leq -1.
\end{equation}
Also, it seems from the above equation and from \eqref{eq:dcoup}
that we must have $\alpha s < 0$ and there is a minimum value
for the magnetic flux $\alpha$.
It is worthwhile observe that bound states and additional
scattering still remain inclusive when $\varepsilon=0$, i.e., no
quantum deformation case, because the condition $\lambda_m<0$ is
satisfied, as it is evident from \eqref{eq:lambdaj}.
It was shown in
Refs. \cite{JPA.1993.26.7637,PRD.2012.85.041701}.

Now, let us comeback to helicity conservation problem.
In fact, the failure of helicity conservation expressed in
Eq. \eqref{eq:scattamp}, it stems from the fact that the
$\delta$ function singularity make the Hamiltonian and the
helicity non self-adjoint operators
\cite{PRD.1977.15.2287,PRL.1983.50.464,PLB.1993.298.63,
NPB.1994.419.323}, hence their commutation must be analyzed
carefully by considering first the correspondent self-adjoint
extensions and after that compute the commutation relation, as
we explain below.
By expressing the helicity operator in terms of the variables
used in \eqref{eq:wavef}, we attain
\begin{equation}
  \hat{h} =
  \left(
    \begin{array}{cc}
      0
      & \displaystyle -i\left(\partial_r
        +\frac{s|m+\alpha|+1}{r}\right) \\
     \displaystyle -i\left(\partial_r
       -\frac{s |m+\alpha|}{r}\right) & 0
    \end{array}
  \right).
\end{equation}
This operator suffers from the same disease as the Hamiltonian
operator in the interval $|m+\alpha|<1$, i.e., it is not
self-adjoint \cite{PRD.1994.49.2092,JPA.2001.34.8859}.
Despite that on a finite interval $[0,L]$, $\hat{h}$  is a
self-adjoint operator with domain in the functions satisfying
$\xi(L)=e^{i\theta}\xi{(0)}$, it does not admit a self-adjoint
extension on the interval $[0,\infty)$ \cite{AJP.2001.69.322},
and consequently it can be not conserved, thus the helicity
conservation is broken due to the presence of the singularity at
the origin \cite{PRD.1977.16.1815,PRL.1983.50.464}.

\section{Conclusion}
\label{sec:conclusion}

We have studied the AB problem within the framework of
$\kappa$-deformed Schr\"{o}dinger-Pauli equation.
The new contribution to the Pauli's term is used to impose a
upper bound in the deformation parameter, 
$\varepsilon < 2.27\times 10^{-9} \;(eV)^{-1}$.
It has been shown that there is an additional scattering for any
value of the self-adjoint extension parameter and for negative
values there is non-zero energy bound states.
On the other hand, the scattering amplitude show a energy
dependency, it stems from the fact that the helicity operator
and the Hamiltonian do not to commute.
These results could be compared with those obtained  in Ref.
\cite{JPA.1993.26.7637} where a relation between the
self-adjoint extension parameter and the gyromagnetic ratio $g$
was obtained.
The usual Schr\"{o}dinger-Pauli equation with $g=2$ is
supersymmetric \cite{PRD.1984.29.2375} and consequently it
admits zero energy bound states \cite{PRA.1979.19.2461}.
However, in the $\kappa$-deformed Schr\"{o}dinger-Pauli equation
$g \neq 2$ and supersymmetry is broken, giving rise to non-zero
energy bound states.
Changes in the helicity in a magnetic field represent a measure
of the departure of the gyromagnetic ratio of the electron or
muon from the Dirac value of $2e/2M$ \cite{Book.1967.Sakurai}.
Hence, the helicity nonconservation is related to nonvanishing
value of $g-2$.

\section*{Acknowledgments}
The authors would like to thank R. Casana and M. M. Ferreira Jr.
for critical reading the manuscript and helpful discussions.
E. O. Silva acknowledges research grants by CNPq-(Universal)
project No. 484959/2011-5.

\bibliographystyle{model1a-num-names}
%\bibliography{abd-scatt}

\end{document}